\documentclass[apsrev4-1,preprint,prd,showpacs,floatfix,superscriptaddress]{revtex4-1}
\usepackage{epsf,color,colordvi}
\usepackage{caption}
\usepackage{graphicx}
\usepackage{amsmath}
\usepackage{times}
\usepackage{subfig}
\usepackage{ulem}  % This package redefines \emph{} to underline
\usepackage{placeins}

\begin{document}

\title{Schottky Anomaly and Hadronic Spectrum}

\author{Aritra Biswas}
\email{aritrab@imsc.res.in}
\affiliation{The Institute of Mathematical Sciences,  Chennai 600113, India}
\author{M V N Murthy}
\email{murthy@imsc.res.in}
\affiliation{The Institute of Mathematical Sciences,  Chennai 600113, India}
\author{ Nita Sinha}
\email{nita@imsc.res.in}
\affiliation{The Institute of Mathematical Sciences,  Chennai 600113, India}

\begin{abstract}

We show that the hadronic ``heat capacity" calculated as a function of 
temperature may be used to infer the possible presence of different 
scales underlying the dynamical structure of hadronic resonances using the 
phenomenon of Schottky anomaly. We first demonstrate this possibility 
with well known meson spectrum in various channels and comment on the 
possibility of using this method as a diagnostic to distinguish the 
exotic states. 

\end{abstract}

%\pacs{}

\date{\today} 

\maketitle

\section{Introduction}

The recent announcement of the discovery~\cite{pentaquark} of the so called penta-quark 
states has rekindled interest in the possibility of exotic hadron 
states. In the last decade, a large number of exotic
mesonic states known as the X, Y, Z states~\cite{Olsen:2014} have been observed by different
experimental collaborations, BELLE, BaBar, BESIII, CDF, CLEO, LHCb etc.~\cite{olsen_ref}.
Long ago, a proposal was made to identify the $\Lambda(1405)$ baryon as a possible 
molecular state of a colourless baryon and meson~\cite{rajaji1, Dalitz, rajaji2}. 
Recently this proposal was confirmed through lattice calculations~\cite{lattice}.

While the existence of such exotic states has been theoretically studied 
over decades~\cite{exoticsth1, bhaduri3, Godfrey:1998pd, exoticsth2, exoticsth3, Godfrey:olsen} 
using models of quark confinement with 
hyperfine interaction, we explore the possibility of identifying states 
which differ in their underlying dynamics due to different interaction scales 
responsible for forming composite states such as mesons. This 
is a model independent analysis which depends entirely on the information 
already contained in the experimental data on the spectra of composite states.

The method itself is not new. It has been in vogue in the study of 
semiconductors with impurities (which may have closely spaced electronic 
spectra) or in the analysis of spectra of deformed nuclear states for a 
long time. In principle when the number of states are large, the high 
temperature behaviour of the specific heat directly yields the 
information about the relevant degrees of freedom in the spectrum- this 
is the well known Dulong-Petit Law. This has been effectively put to 
use in the analysis of the light baryon spectrum by Bhaduri and 
Dey~\cite{bhaduri1} where they have shown that even with truncated 
spectrum of light quark baryon states, the degrees of freedom of the 
system may be inferred by comparing the models with the experimental 
spectrum through the so called Schottky peak. However, such an analysis 
cannot be applied to the meson spectrum to infer the degrees of freedom
as may be inferred from an analysis of the meson spectrum.

On the other hand, when a truncated spectrum is available with no 
saturation possible, the specific heat at low temperature displays a 
Schottky peak (or peaks) which is an indication of the relevant scale 
(scales) in the system. Therefore the method we use here, consists of 
analyzing the heat capacity $C_V$ of a spectrum of states which may 
contain finitely many states that may not lead to saturation of the 
specific heat. The existence of Schottky peak (peaks) is taken as an 
indication of the presence of an interaction scale (scales) and analyzed 
further. This provides a possible model independent diagnostic of the 
presence of unusual or exotic states. The only input used is the
experimentally measured spectrum of states. 

In sec.II, we illustrate the method with a set of assumed ideal spectra 
closely following the illustrations adopted from Ref.~\cite{bhaduri2}. 
Though the method has been widely used in other fields, it may not be 
familiar to the practitioners in particle physics phenomenology. In 
sec.III we first discuss the charmonium spectrum to illustrate the 
applicability of the method to the measured spectrum of states using the 
lists provided in PDG~\cite{pdg}. Charmonium spectrum provides a template
for the analysis of other states. Further, we discuss other cases including the 
bottomonium and open-charm/bottom mesons and show that interesting facts emerge by a simple 
application of the idea of the Schottky anomaly. We conclude with a mention
of caveats and how the method may be fruitfully used as and when more data
becomes available particularly for the conjectured exotic bottomonium states.

\section{Heat capacity of an ideal system}

In order to illustrate the method, consider the simplest case of a two 
level system with an energy gap given by $\Delta$. The energy gap is an 
indication of the scale in the Hamiltonian. The canonical partition 
function of the system is simply given by
\begin{equation}
Z=1+e^{-\beta \Delta},
\label{idealz}
\end{equation}
where $\beta=1/k_B T$ is the inverse temperature. Hereafter we set the 
Boltzmann constant $k_B=1$ and the temperature is given in energy units.
For the purpose of illustration here we have not assumed any occupancy factors. 
The temperature is introduced here simply as a mathematical parameter 
to define the partition function of the system and no assumption is made 
regarding the system being in a heat bath in equilibrium. It is a parameter that 
is used to calculate the specific heat or more precisely energy 
fluctuations. 

The specific heat of the system at constant volume may be defined as
\begin{equation}
C_V=\beta^2\left[\frac{1}{Z}\frac{\partial^2 Z}{\partial\beta^2}
-\left(\frac{1}{Z}\frac{\partial Z}{\partial\beta}\right)^2\right]
=\beta^2[\langle E^2\rangle-\langle E\rangle^2].
\label{cvdef}
\end{equation}  
In systems with constant density, we may replace $C_V$ by $C_V$. Substituting
for the partition function of the two level system given in eq.(\ref{idealz})
we have
\begin{equation}
C_V= \beta^2 \frac{\Delta^2 e^{-\beta\Delta}}{(1+e^{-\beta\Delta})^2}.
\end{equation}
When $C_V$ is plotted against $\beta\Delta$ the Schottky peak appears at 
a value $\beta\Delta\approx 2.4$ with an exponential tail at higher 
values of $\beta$. The location of the peak is a function of the energy 
gap in the system. In general the Schottky peak occurs in systems with 
few energy levels where gaps are a result of a single scale parameter. 
If however, there is more than one independent scale parameter 
responsible for the energy levels, then peaks appear whenever the 
temperature is sufficient to cross the gap signaling a change in the 
entropy of the system. At high temperatures when all the levels are 
equally possible, a plateau appears signaling very little change in the entropy.
 
\begin{figure}[ht]
\centering
\includegraphics[width=0.8\textwidth]{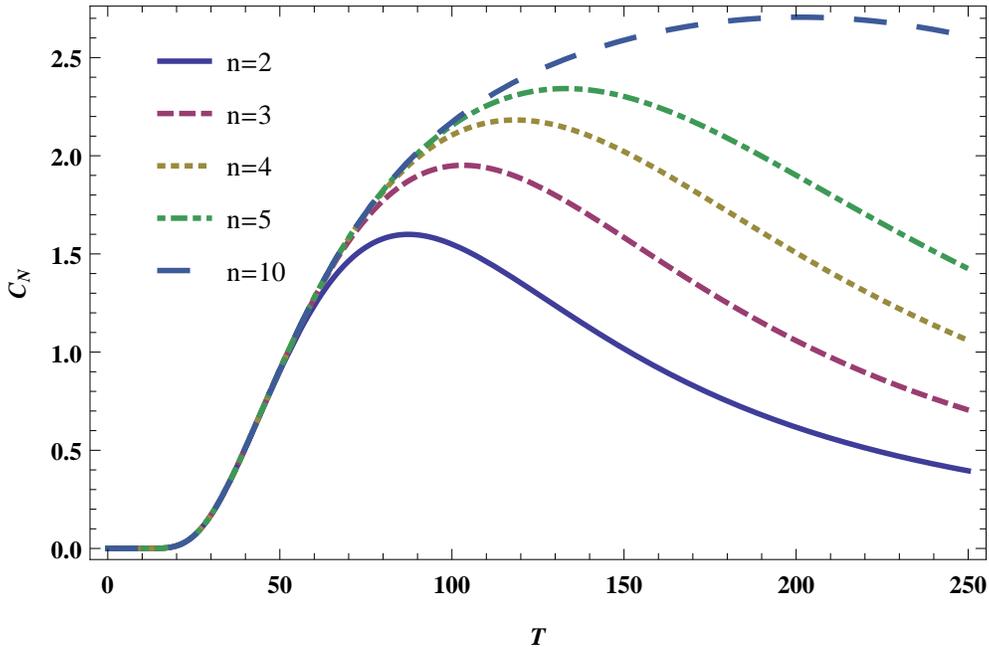}
\caption{Schematic illustration of the Schottky peak in the 3-dimensional
harmonic oscillator spectrum showing the effect of truncation of 
the spectrum.}
\label{fig1}
\end{figure}

To further expand on this theme, we next consider an ideal spectrum, namely 
the spectrum of a three-dimensional Harmonic oscillator. 
The partition function of the system is given by
\begin{equation}
Z_1=\sum_{n=0}^{\infty}D(n)e^{-\beta\hbar\omega(n+3/2)},
\label{idealho}
\end{equation}
where the oscillator parameter $\omega$ defines the scale in the 
problem and $D(n)$ is the degeneracy of the level. 
In Fig.~\ref{fig1}, we show the effect of truncation on the 
specific heat when plotted as a function of $T=1/\beta$ with 
$\hbar\omega=200~MeV $ (say). As we include more and more orbitals, the
specific heat tends to reach the required saturation, while the peak persists
in the truncated spectrum.  
As we shall see later, in any realistic hadronic spectrum, especially in the 
heavy quark sector, we do not need more
than 2 to 3 orbitals to count the observed spectrum of states. For these cases the location of 
the Schottky peak is close to $\beta\hbar\omega\approx2.42$. 

\begin{figure}[ht]
\centering
\subfloat[][]{\includegraphics[width=0.49\textwidth]{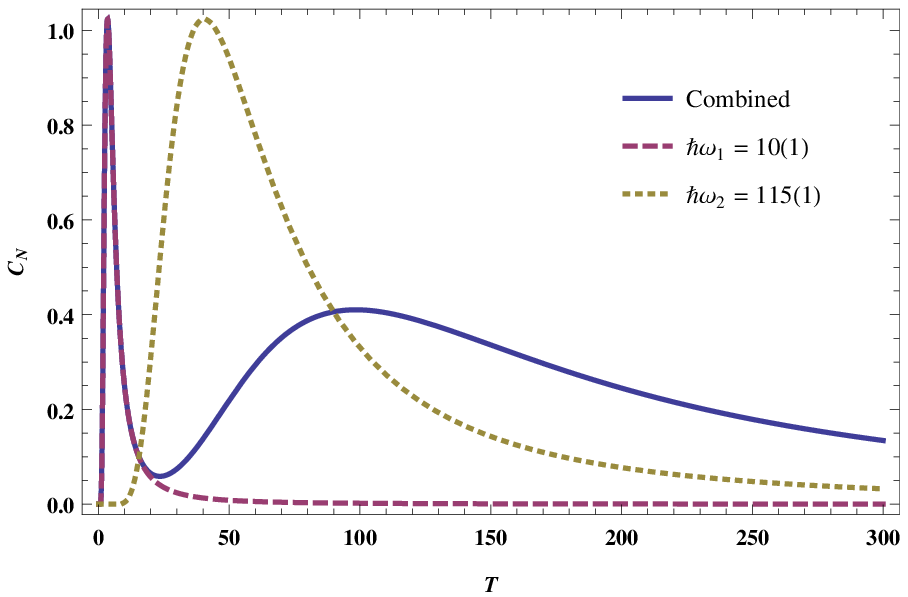}}\hfill
\subfloat[][]{\includegraphics[width=0.49\textwidth]{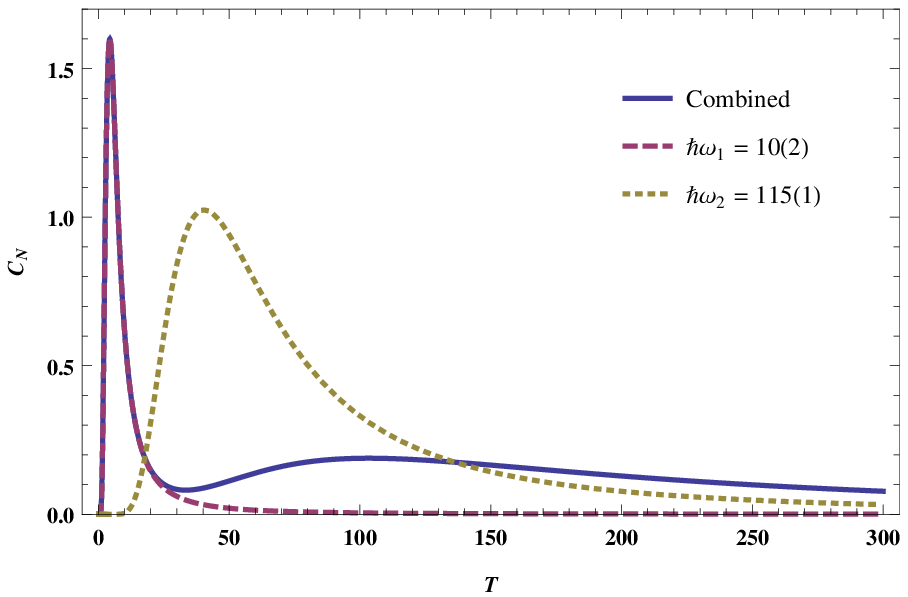}}\\
\subfloat[][]{\includegraphics[width=0.49\textwidth]{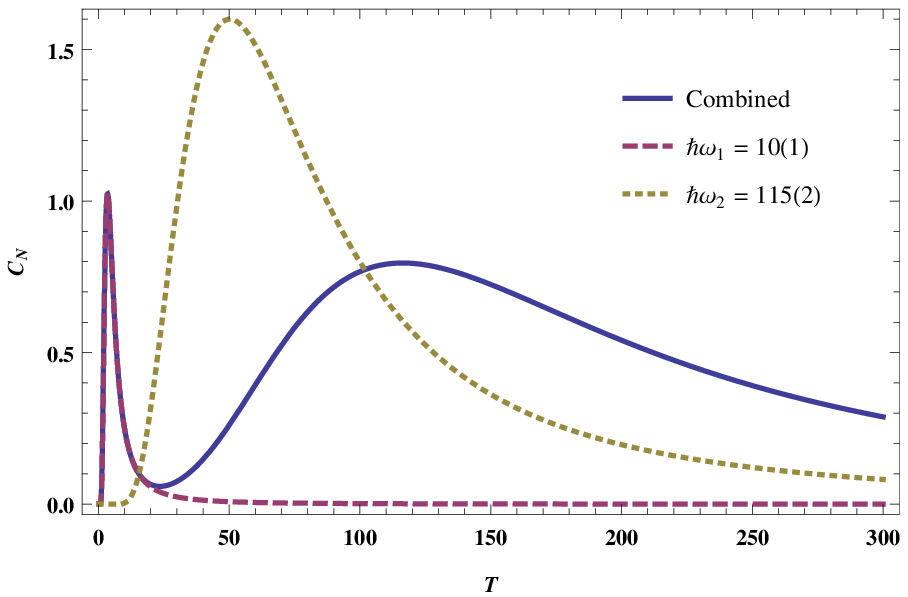}}\hfill
\subfloat[][]{\includegraphics[width=0.49\textwidth]{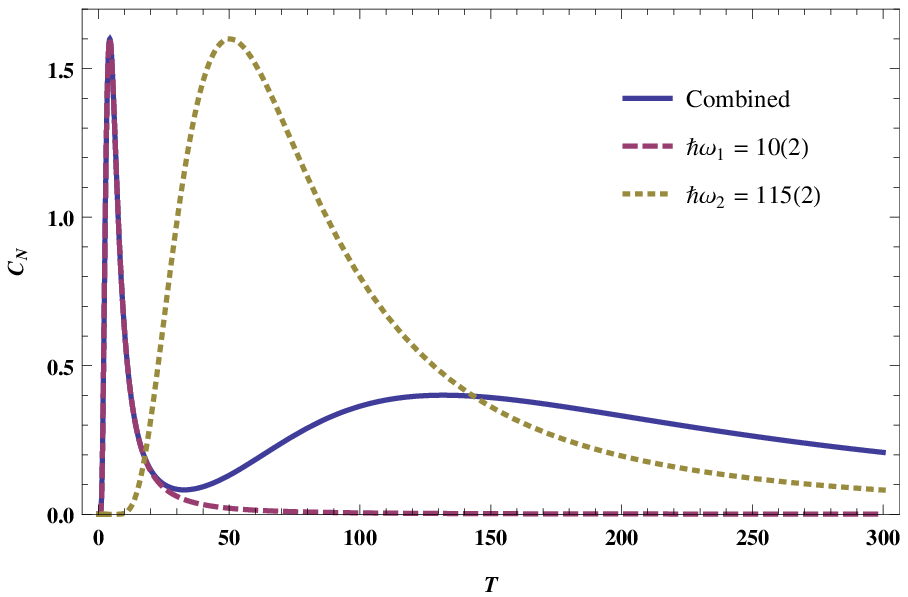}}
\caption{Schematic illustration of the Schottky peaks in the ideal case when the data
contains two scales. The individual Schottky peaks corresponding to separate spectra (harmonic oscillator
spectrum), as well as the combined one (solid line), are shown. The frequencies 
are the same for all of the four figures as shown ($\hbar\omega_1=10$ MeV and $\hbar\omega_2=115$ MeV).
The cutoff in the principle quantum number $n$ (indicated in brackets) are: (a) $n_1=n_2=1$ (b) $n_1=2$
$n_2=1$ (c) $n_1=1$ $n_2=2$ and (d) $n_1=n_2=2$.}
\label{fig2}
\end{figure}

In order to simulate a realistic spectra where there is a possibility of 
more than one energy scale in operation, we combine the spectra of 
two such systems which differ in $\hbar\omega$ significantly. In Fig.~\ref{fig2}
we show the individual spectra separately as well as the spectra for the combined single
set. The effect of combining is to normalize the specific heat
with a single partition function given by,
\begin{equation}
Z(\beta)= Z_1(\beta,\hbar\omega_1)+Z_2(\beta,\hbar\omega_2).
\end{equation}
This is somewhat artificial but nevertheless we use this for the purpose
of illustrating the effect of the presence of multiple scales in the spectra
when the states are combined in a listing in the absence of any dynamical
information.

As can be seen  from Fig.~\ref{fig2} the amplitude and location of the
two separate Schottky peaks depend on truncation apart from the oscillator
frequencies. When combined, the peak corresponding to the lower
frequency remains unchanged while the one corresponding to higher frequency
is sensitive to truncation. This is simply due to the domination of the lower
scale in the partition function.

We exploit this sensitivity to different scales in the problem, and apply it in the 
analysis of experimental spectra of sets of mesons in the next section. 
Different scales may arise from different terms in the same Hamiltonian, 
for example the confinement scale could be very different from the 
splitting of states with different spins, namely the hyperfine splitting. 
Alternatively the data set could contain 
states which may arise from different underlying dynamics 
and hence have different scales. Often it may so happen that the data set is 
missing a certain number of states which also introduces an artificial 
scale (or energy gap) in the analysis. Nevertheless, the analysis of 
Schottky peaks in the experimental spectra may reveal some hidden 
scales on the average. We must however caution that this analysis will not reveal any 
detailed dynamics (or the system Hamiltonian) underlying the formation 
of the bound states. 

\section{Analysis of the experimental data}

We consider an analysis, based on the template given in the previous 
section, of the experimental data on mesons. We divide the data sets 
according to their flavours and start from the simplest cases, where 
there is no complication arising from the isospin degeneracies. In this
analysis we only use the identification given in the particle data group
tables without any further theoretical  bias. We focus on the heavy quark
sector.

\subsection{The spectra of charmonium states}

\begin{figure}[ht]
\centering
\subfloat[][]{\includegraphics[width=0.49\linewidth]{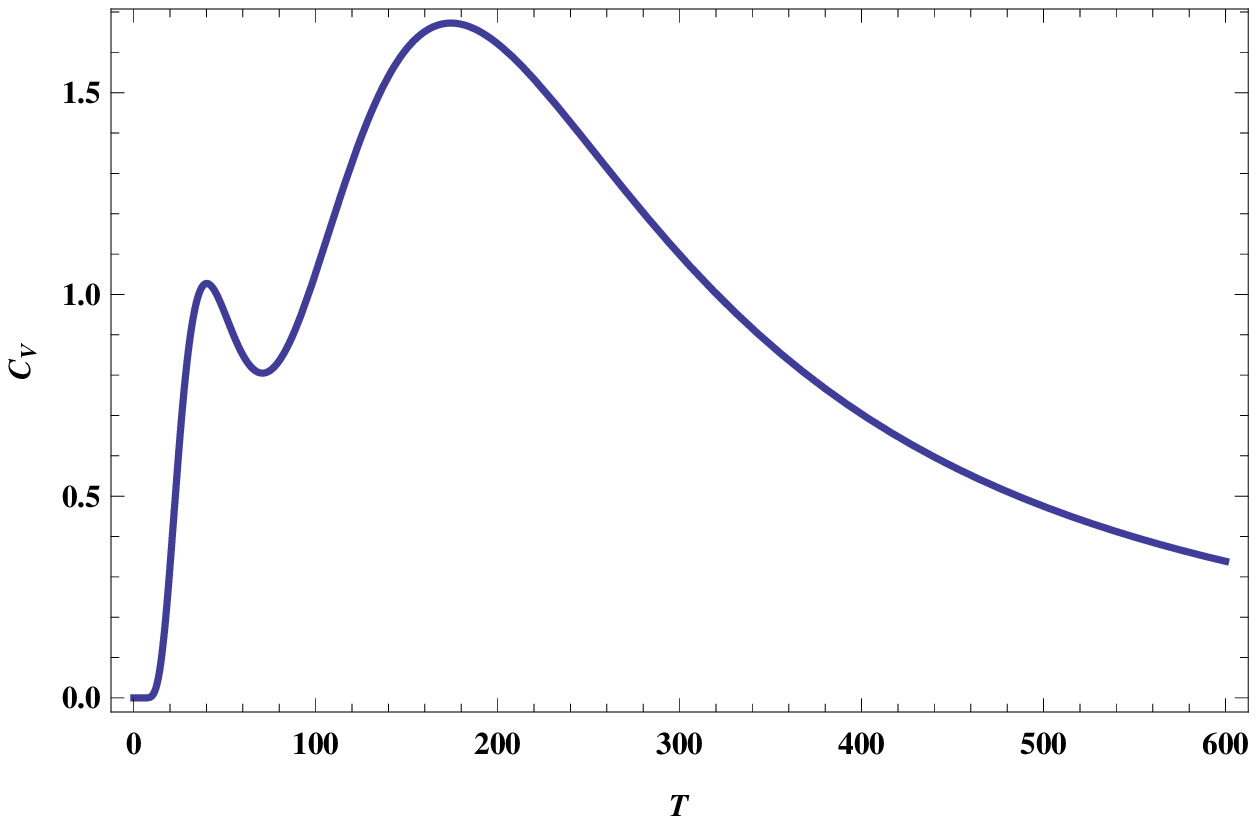}\label{fig31}}\hfill
\subfloat[][]{\includegraphics[width=0.49\linewidth]{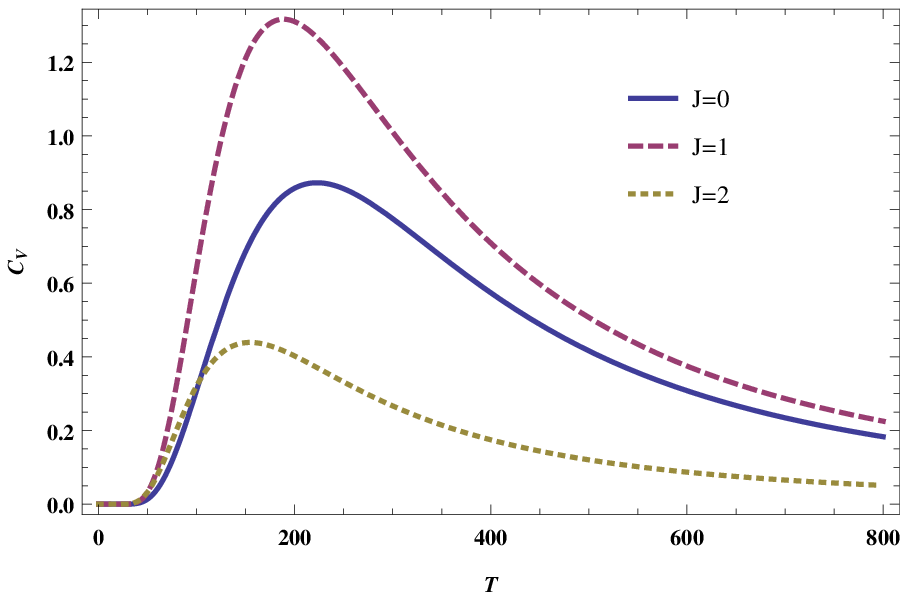}\label{fig33}}
\caption{The specific heat of charmonium states plotted as a function of temperature 
(expressed in MeV units) with (a) all the states taken together (i.e. all the states in Table~\ref{tab1})
and (b)after separating them according to their spins $J=0,1,2$ (columns 3, 6 and 9 of Table~\ref{tab1}).} 
\end{figure}

We first
choose to analyze the spectra of the $c\bar{c}$ (charmonium) states as identified in the
PDG~\cite{pdg}, listed in Table~\ref{tab1}.
\begin{table}[ht]
\captionsetup{justification=raggedright,
singlelinecheck=false
}
\centering
\begin{tabular}{|c|c|c|c|c|c|}\hline
~$J=0$ States~& Mass &~$J=1$ States~& Mass &~$J=2$ States~& Mass \\
\hline\hline 
$\eta_c(1S)$   & 2983.6  &$J/\psi(1S)$   & 3096.92 &$\chi_{c2}(1P)$& 3556.2\\\hline
$\chi_{c0}(1P)$& 3414.75 &$\chi_{c1}(1P)$& 3510.66 &$\chi_{c2}(2P)$& 3927.2\\\hline
$\eta_c(2S)$   & 3639.4  &$h_c(1P)$      & 3525.38 &               & \\\hline
$\chi_{c0}(2P)$& 3918.4  &$\psi(2S)$     & 3686.11 &               & \\\hline
               &         &$\psi(3770)$   & 3773.15 &               & \\\hline
               &         &$\psi(4040)$   & 4039.6  &               & \\\hline
               &         &$\psi(4160)$   & 4191    &               & \\\hline
               &         &$\psi(4415)$   & 4421    &               & \\\hline
\hline
\end{tabular}
\caption{Charmonium masses given in MeV along with their total angular momentum $J$. Other
quantum numbers are not needed for this analysis. }
\label{tab1}
\end{table}
Since the 
number of resonant states is not too small (but no where near saturation), the 
spectra may be used to determine the scales involved in the problem. 
Since these states involve heavy quarks, one may also 
compare the results of the analysis with models of confinement if 
necessary. In Fig.~\ref{fig31} we show the specific heat of the 
$c\bar{c}$ spectrum.

When all the $c\bar{c}$ are taken together, the spectrum shows 
two clear peaks at $T\approx 40$ MeV and $T\approx 190$ MeV 
indicating the existence of two well defined scales in the spectrum of 
states. This is not surprising since all of the 
states with different $J$ values, $J=0,1,2$ are included. 
As a result not only the confinement scale, but 
also the hyperfine(HF) splitting scale comes into operation. We may think of 
the peak at $40$ MeV as being due to the HF splitting whereas the peak at 
190 MeV resulting from the confinement scale. For a two level system, using 
the relation $\beta\Delta=2.4$, the corresponding energy gap turns out 
to be $96$ MeV for hyperfine splitting and $450$ MeV for the confinement 
potential.  These are reasonable values from the point of view of quark 
models though we need not assume any particular model for the charmonium 
states. These scales reflect the average behaviour, since the actual value of HF
splitting depends on the orbital in which it is calculated. For example, in 
the ground state the HF splitting is of the order of $100$ MeV. 

These two scales must correspond to the contribution to the masses of the resonances from the 
QCD inspired potential models (or in Lattice calculations) where the central potential is assumed to 
be a combination of the linear confinement potential as well as the 
Coulomb like potential arising from one gluon exchange interaction: 
\begin{equation} 
V(r) = C r +\frac{4\alpha_s}{3r}, \label{central} 
\end{equation} 
while the hyperfine interaction, which also arises along 
with the Coulomb interaction from the one-gluon exchange interaction, in 
general has the form, 
\begin{equation}
V_{hf}(r)= 
k\frac{\sigma_1.\sigma_2}{m_1 m_2}f(r). \label{central} 
\end{equation}

Such a non-relativistic reduction of the QCD inspired potential is a
good approximation in the heavy quark sector. In this paper we do not 
go into the details of the models but give these here only to indicate
possible origins of the Schottky peaks. It is sufficient to point out
that the central potential which is spin and flavour 
independent, contributes in all sectors while the HF splitting 
depends inversely on the quark masses as well as the particular
orbital in which the spin dependent potential is evaluated.

To clarify this notion further, in Fig.~\ref{fig33} we plot the states
corresponding to different $J$ values separately.
It is easily seen that the peak corresponding to HF splitting 
disappears as it should when different $J$ states are plotted separately.
As a result there is only one scale in the problem due to the central  
potential. The slight shift in the peaks in the specific heat is due to 
the effect of truncation of the states as noticed in the ideal case in 
the previous section. While there are more states in $J=0,1$, there are 
only two states in the case of $J=2$. In either case, the number of 
orbitals involved will not be more than two to three even if a model is 
invoked.

\begin{table}[ht]
\centering
\captionsetup{justification=raggedright,
singlelinecheck=false
}
\begin{tabular}{|c|c|c|c|c|c|}\hline
~$J=0$ States~& Mass &~$J=1$ States~& Mass  \\
\hline\hline 
$X(3940)$ \cite{exoticsexp2}& 3942  &$X(3872)$ \cite{exoticsexp1,lhcb1}       & 3871.69 \\\hline
$Y(4140)$ \cite{exoticsexp4}& 4144  &$X(3900)$ \cite{exoticsexp12}            & 3896.35 \\\hline
$X(4160)$ \cite{exoticsexp5}& 4156  &$X(4260)$ \cite{exoticsexp3,exoticsexp6} & 4251    \\\hline
                            &       &$Y(4360)$ \cite{exoticsexp9}             & 4361    \\\hline
                            &       &$X(4430)^{\pm}$ \cite{exoticsexp16}      & 4485    \\\hline
                            &       &$X(4660)$ \cite{exoticsexp11}            & 4664    \\\hline
                            &       &$Y(4008)$ \cite{exoticsexp3}             & 4008    \\\hline
                            &       &$Z_c^+(4020)$ \cite{exoticsexp13}        & 4024    \\\hline
                            &       &$Z_1^+(4050)$ \cite{exoticsexp14}        & 4051    \\\hline
                            &       &$Z^+(4200)$ \cite{exoticsexp15}          & 4196    \\\hline
                            &       &$Z_2^+(4250)$ \cite{exoticsexp14}        & 4248    \\\hline
                            &       &$X(4630)$ \cite{exoticsexp10}            & 4634    \\\hline
\hline
\end{tabular}
\caption{The exotics in the charmonium mass range. All masses except that of $Y(4008)$
are taken from PDG~\cite{pdg}. The mass of $Y(4008)$ as well as some of the $J$ assignments not given in 
PDG are from~\cite{Olsen:2014}.}
\label{tab_cc_ex}
\end{table}

\begin{figure}[ht]
\centering
\subfloat[][]{\includegraphics[width=0.49\linewidth]{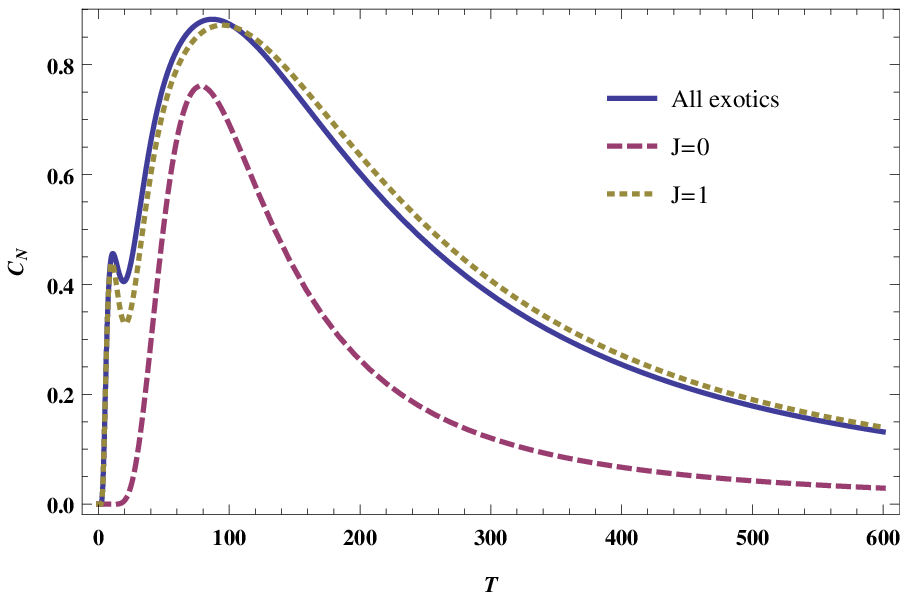}\label{ccbar_ex_J_sep}}\hfill
\subfloat[][]{\includegraphics[width=0.49\linewidth]{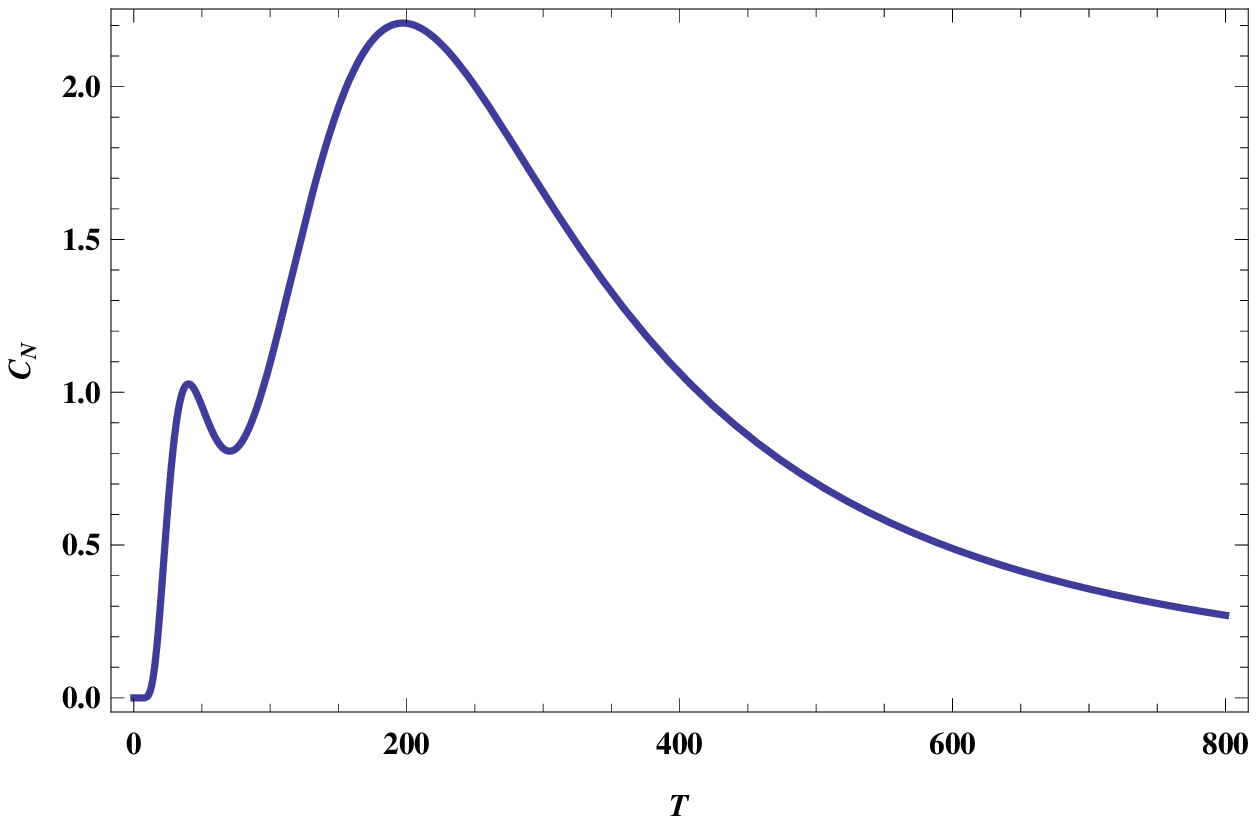}\label{ccbar_with_ex}}
\caption{The specific heat of the exotics as listed in~\cite{Olsen:2014} and~\cite{pdg}
in the charmonium region, plotted as a function of temperature 
(a) for the pure exotics spectrum (i.e. states
from Table~\ref{tab_cc_ex}) and the same separated according to the corresponding $J$'s  only and 
(b) combined with all the other $c\bar{c}$ states (i.e. all states from Tables~\ref{tab1}
 and~\ref{tab_cc_ex} combined)}
\end{figure}

Thus, having understood the charmonium spectrum where the two scales, confinement
and HF interaction energies stand out, we may now apply the method to the so-called 
exotic states listed along with the charmonium
states. Table~\ref{tab_cc_ex} shows the states used in the present analysis.
Some of these assignments may be tentative or educated guesses as discussed in~\cite{Olsen:2014}.
Nevertheless the specific heat of these states may be calculated and is
shown in Fig.~\ref{ccbar_ex_J_sep}.

The specific heat plot shows two peaks, one reasonably broad peak close to 100 MeV 
and a very sharp one below 10 MeV. The peak at 100 MeV is an unique characteristic of
only the exotics. For a two level system, using 
 $\beta\Delta=2.4$, the corresponding energy gap is found to be $240$ MeV 
 in contrast to the pure $c\bar{c}$ 
case, where it is close to 450 MeV, and can be attributed to the confinement
scale. The $\textquoteleft$confinement peak' is also present in the combined plots for the charmonium
states along with the exotics (Fig.~\ref{ccbar_with_ex}), but not for the exotics alone.
This indicates the presence of a scale that is different from the confinement, hyperfine
or any other scale that is exhibited by the established charmonium states. This definitely
hints at a different interaction mechanism for formation of the exotic meson states. 
This may be called the $\textquoteleft$exotic' scale, as, in the next section we show that this same
scale is present for the exotics in the bottomonium mass range as well.

\begin{figure}[ht]
\centering
\subfloat[][]{\includegraphics[width=0.49\linewidth]{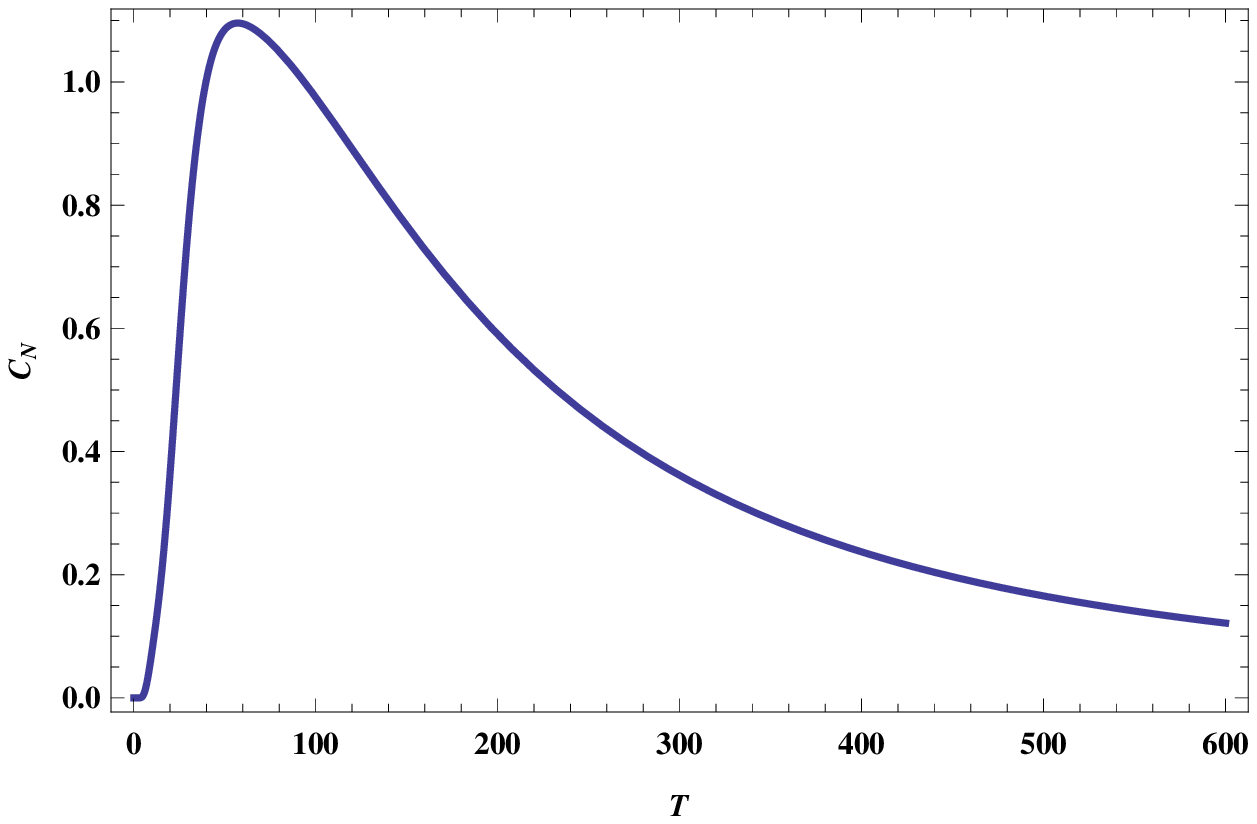}\label{ccbarexnoX}}\hfill
\subfloat[][]{\includegraphics[width=0.49\linewidth]{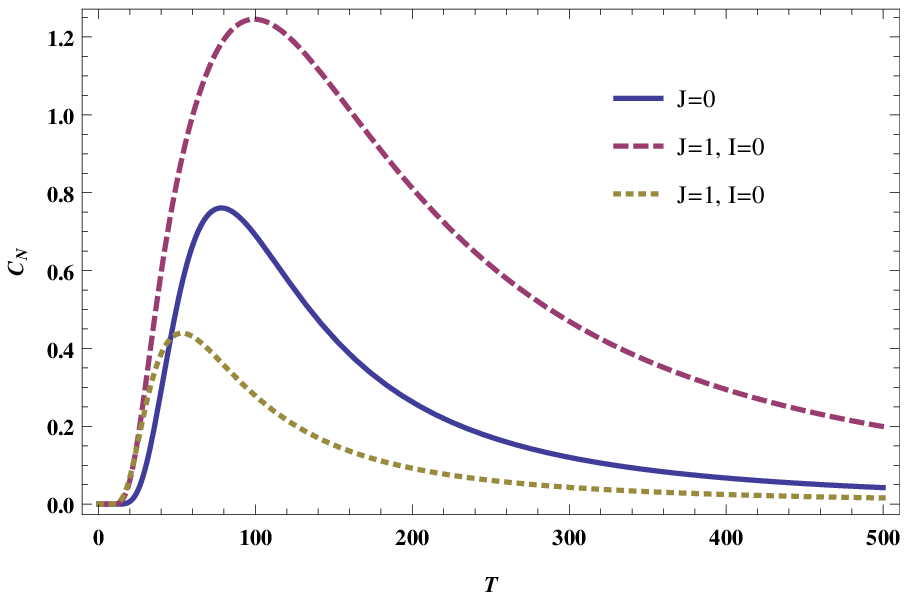}\label{ccbar_ex_JI_sep}}
\caption{(a) The single peak obtained by omitting the resonance at 3872 MeV from the other exotics, 
(b) The exotics plotted after the assignment of the $I$ values, separated into corresponding 
groups with same $J$ and $I$}
\end{figure}

It is easily seen that the sharp peak below 10 MeV arises 
due to contributions from the first two states which are indeed very close 
in energy. Since the mass of $X(3872)$ is very close to the $D^0{\bar{D}}^{*0}$
threshold, it was described in many reports earlier 
as a molecular state~\cite{X38721,X38722,X38723,X38724,X38725}. 
However, the characteristics of X(3872) production in high energy $p\bar{p}$ and $pp$ collisions, as
reported by LHCb~\cite{lhcb1} and CMS~\cite{CMS} and discussed in~\cite{X38726}, match those of a tightly
bound state, rather than that of a molecule.
If this lowest state is removed from the list of exotics 
it is seen that the sharp peak below 10 MeV disappears leaving a single peak 
close to 60 MeV as shown in Fig.~\ref{ccbarexnoX}. The nature of the state at $3872$ MeV
cannot be inferred from the present analysis. The appearance of more such states, if at all,
is needed to clarify its nature. Nevertheless if it is included in the present set, it possibly 
indicates the existence of states with much weaker interaction.

There can be a second explanation for the sharp peak below 10 MeV. The 3896.35 and the 
4024 MeV states have been claimed to be isospin triplets, but not yet confirmed. Assuming them
to be isospin triplets, and hence taking care of the $I$ symmetry along with the $J$ symmetry
(by assuming others to be $I=0$), the plots immediately exhibit the absence of the said peak,
even though the 3872 resonance is included in the $J=1$, $I=0$ set (see Fig.~\ref{ccbar_ex_JI_sep}).
The peaks are close to $T=100$ $(J=0)$, $T=8$0 $(J=1$, $I=0)$ and $T=50$ $(J=1$, $I=1)$
in 
MeV units. These differences may be due to a combination of isospin symmetry breaking as well
as truncation effects.

\subsection{The spectra of bottomonium states}

Encouraged by the considerations in the charmonium spectrum, we now analyze the 
bottomonium spectrum, which includes the states given table~\ref{tab2}. Here
there are fewer observed states.
\begin{table}[ht]
\centering
\captionsetup{justification=raggedright,
singlelinecheck=false
}
\begin{tabular}{|c|c|c|c|c|c|}\hline\hline
~$J=0$ States~& Mass &~$J=1$ States~& Mass &~$J=2$ States~& Mass \\
\hline\hline 
$\eta_b(1S)$   & 9398    &$\varUpsilon(1S)$   & 9460.3  &$\chi_{b2}(1P)$  & 9912.21\\\hline
$\chi_{b0}(1P)$& 9859.44 &$\chi_{b1}(1P)$     & 9892.78 &$\varUpsilon(1D)$& 10163.7\\\hline
$\eta_b(2S)$   & 9999    &$h_b(1P)$           & 9899.3  &$\chi_{b2}(2P)$  & 10268.7\\\hline
$\chi_{b0}(2P)$& 10232.5 &$\varUpsilon(2S)$   & 10023.3 &                 & \\\hline
               &         &$\chi_{b1}(2P)$     & 10255.5 &                 & \\\hline
               &         &$h_b(2P)$           & 10259.8 &                 & \\\hline
               &         &$\varUpsilon(3S)$   & 10355.2 &                 & \\\hline
               &         &$\varUpsilon(4S)$   & 10579.4 &                 & \\\hline
               &         &$\varUpsilon(10860)$& 10876   &                 & \\\hline
               &         &$\varUpsilon(11020)$& 11019   &                 & \\\hline
\hline
\end{tabular}
\caption{Bottomonium masses given in MeV along with their total $J$. Other
quantum numbers are not needed for this analysis.}
\label{tab2}
\end{table}

\begin{figure}[ht]
\centering
\subfloat[][]{\includegraphics[width=0.49\linewidth]{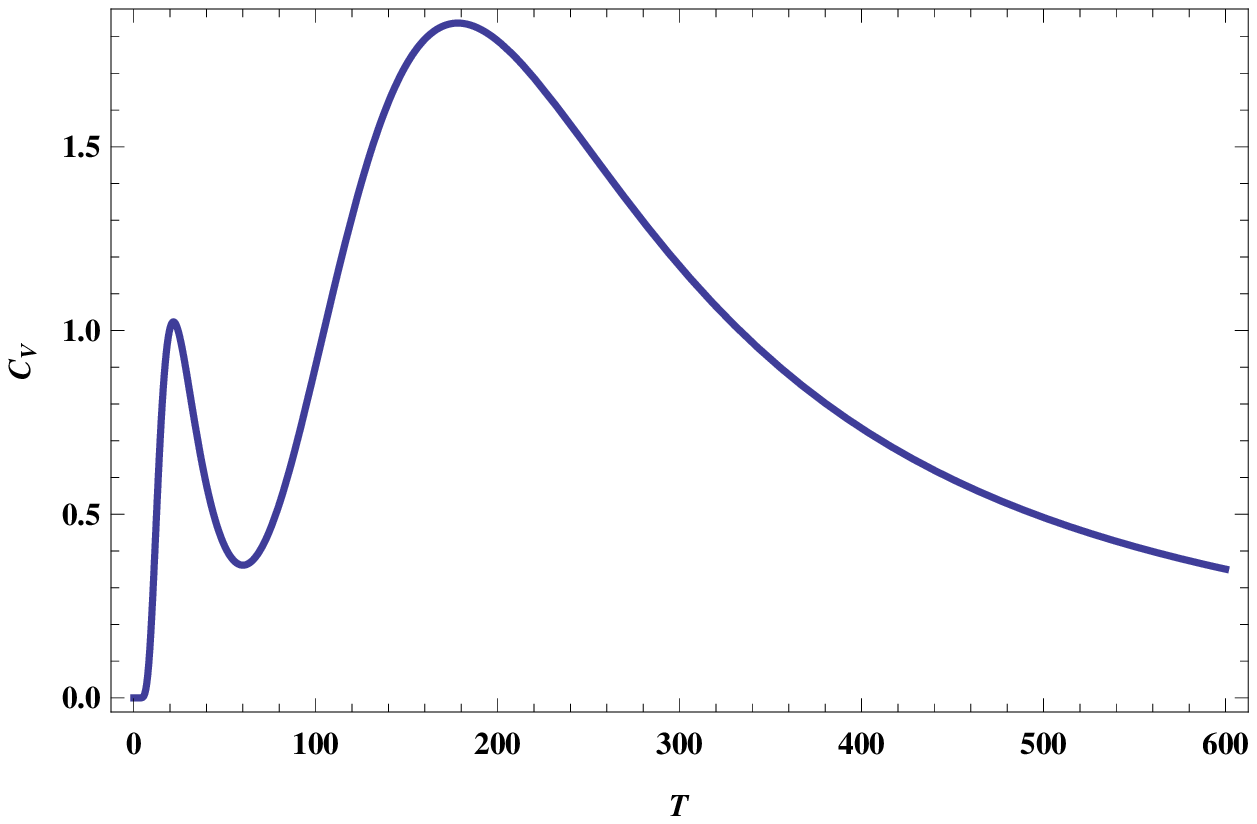}\label{bbbar}}\hfill
\subfloat[][]{\includegraphics[width=0.49\linewidth]{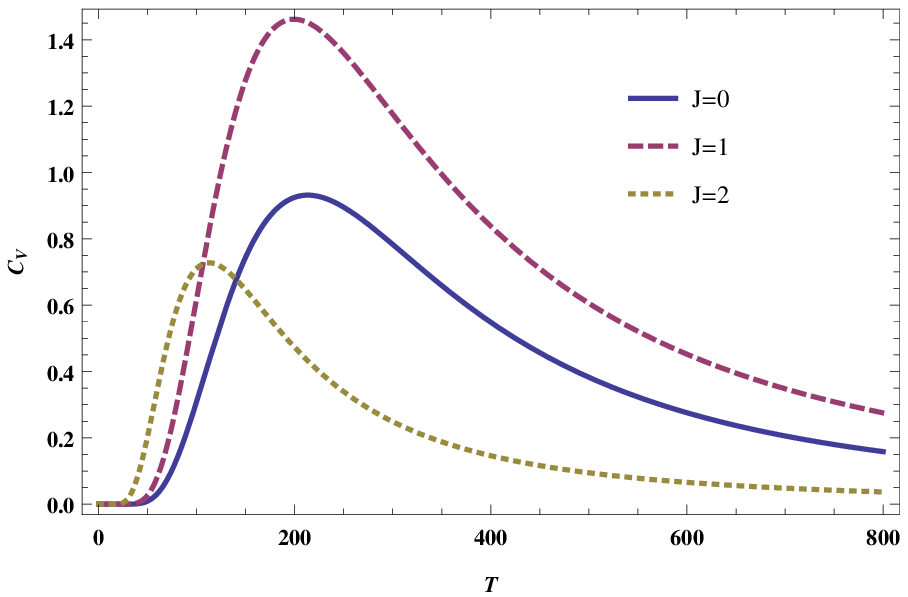}\label{bbbar_J_sep}}
\caption{The specific heat of bottomonium states plotted as a function of temperature 
(expressed in MeV units)
(a) with all the states in Table~\ref{tab2} and
(b) after 
separating them according to their spins $J=0,1,2$.}
\end{figure}

When the specific heats corresponding to these states are calculated and 
plotted together, the situation is quite similar to that of the 
charmonium states. One observes two clear peaks close to $T=185$ MeV 
and $T=22$ MeV as seen in Fig.~\ref{bbbar}. The peak at lower $T$ 
disappears when the states are separated into groups with the same $J$ 
value and plotted again (Fig.~\ref{bbbar_J_sep}). This shows that 
the peak at lower $T$ corresponds to the HF splitting and is thus 
absent when the states are separated according to spin, while that at 
higher $T$ reflects the confinement scale and hence it remains even in 
Fig.~\ref{bbbar_J_sep}. The peak for the J = 2 states is at a slightly lesser T
value, but  then, there are just three corresponding states. With more 
bottomonium states in $J=2$ sector, we expect the peak to 
shift to the right and come close to the $T=185$ MeV value. Interestingly the HF
peak in Fig.~\ref{bbbar} occurs at a reduced temperature as it should, but it does not scale with 
the masses in accordance with Eq.~\ref{central}. The difference may be an effect
arising from the matrix elements. In the s-state, this may be approximated by $|\psi_s(0)|^2$.
It is therefore possible that {\it the bottomonium wave functions are 
more sharply peaked compared to the charmonium states.}

\begin{figure}[ht]
\centering
\subfloat[][]{\includegraphics[width=0.49\linewidth]{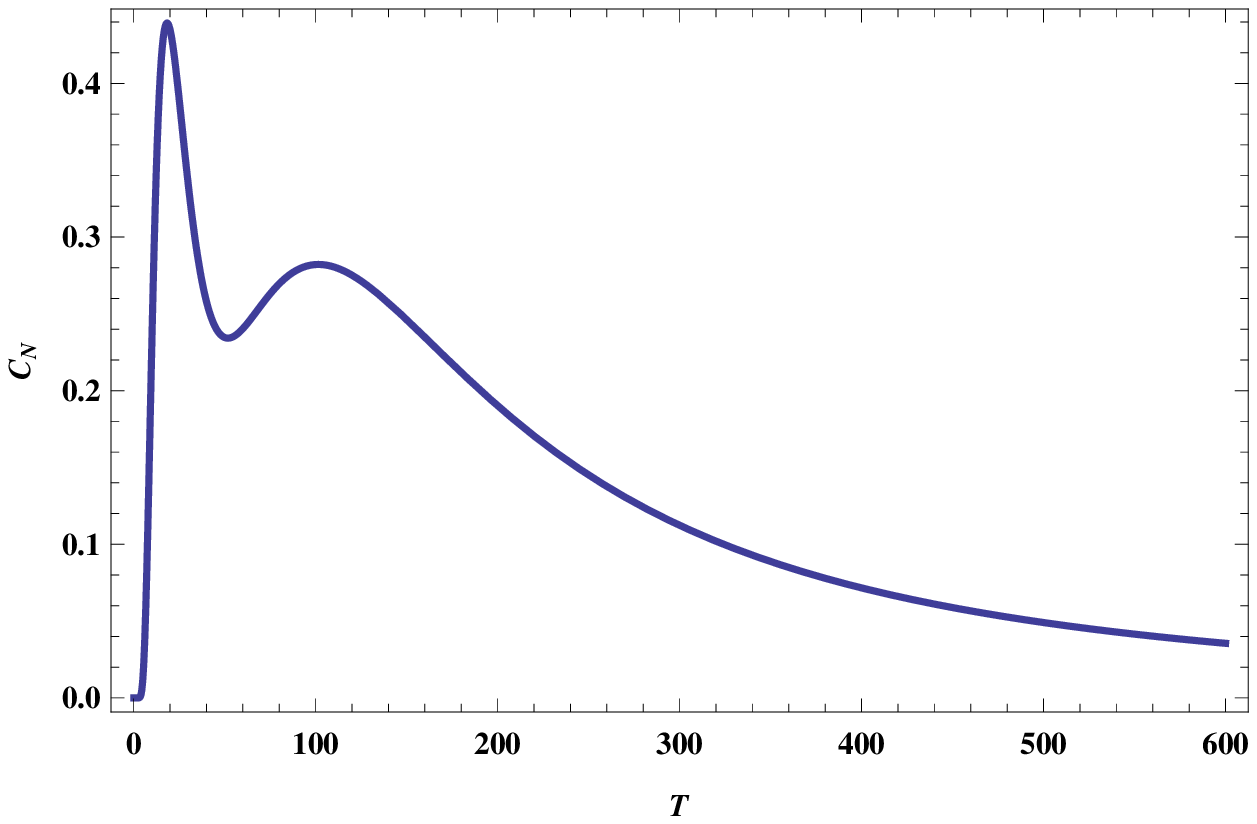}\label{bbbar_exotics}}\hfill
\subfloat[][]{\includegraphics[width=0.49\linewidth]{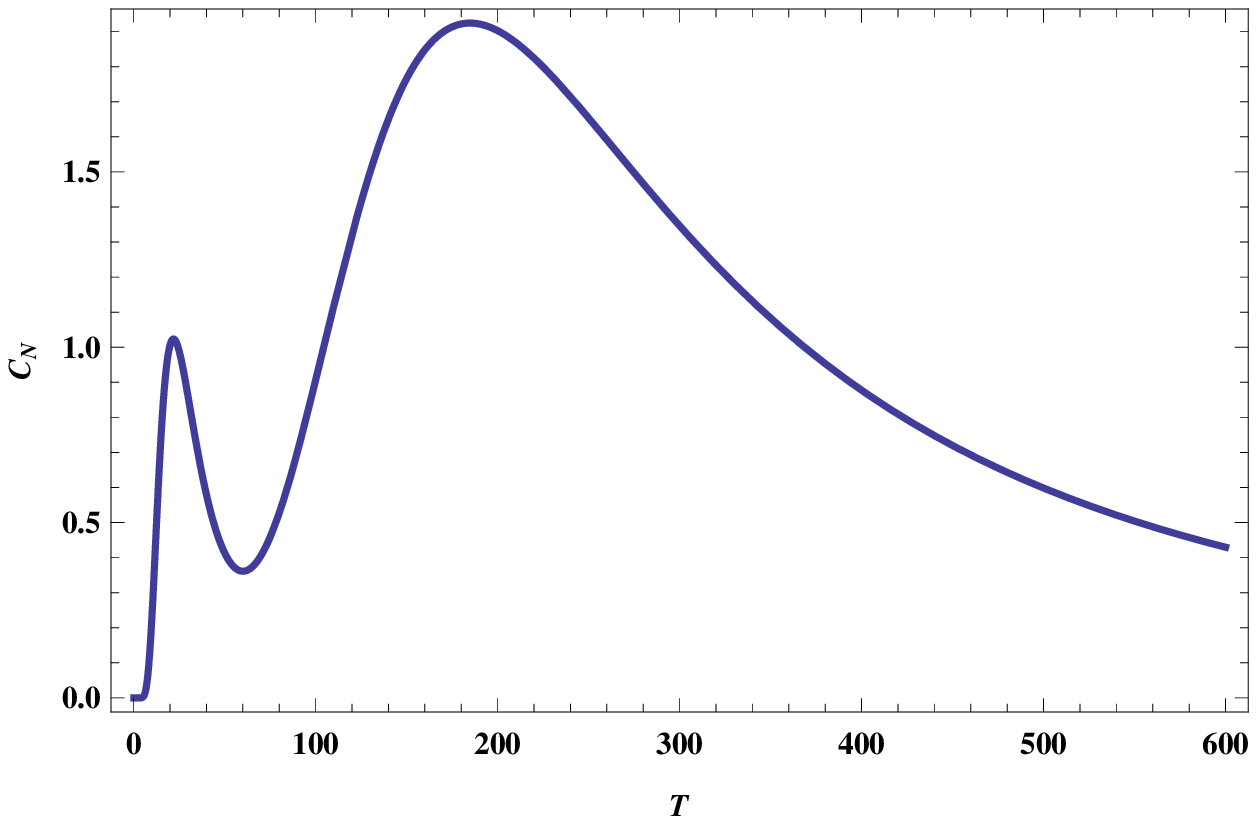}\label{bbbarexall}}
\caption{The specific heat of the exotics as listed in PDG in the 
bottomonium mass range (a) with only the exotic states and (b) combined with pure bottomonium states 
(i.e. combined with all the states of Table~\ref{tab2})} 
\end{figure}

Next, we repeat the analysis with the so-called exotics listed by PDG in 
the bottomonium mass range. The states are $Y_b(10890)$~\cite{exoticsexp17}, 
$X(10650)^{\pm}$~\cite{exoticsexp18} and $X(10610)$~\cite{exoticsexp18,exoticsexp19}
with masses $10888.4$, $10652.2$ and $10608.1$ respectively (all in MeV). 
All of them have $J=1$ and $I=1$.
When plotted against temperature, they again 
exhibit two well defined peaks at approximately $T=19$ and $T=101$ MeV.
(Fig.~\ref{bbbar_exotics}). Since the quantum numbers of these states 
are not well defined, it is difficult to comment on these scales. 
However, similar to the charmonium exotics, the $\textquoteleft$exotic' peak close to 100 MeV 
is present for the bottomoniium exotics also, even with as few as three states.
This leads to the important conclusion that the underlying mechanism may be
similar for both the charmonium and bottomonium exotics. 
The validity of this claim can be tested in future with the advent of more data,
especially in the bottomonium sector.

Similar to their charmonium counterpart, when the exotics are combined together
with established bottomonium states and plotted, the spectra is almost similar to that of 
pure bottomonium states displaying a peak corresponding to the hyperfine splitting and 
another close to $T=200$ MeV corresponding to the confinement scale (Fig.~\ref{bbbarexall}). 

\subsection{The spectra of open charm states}

In the previous two sub-sections, we have analyzed the heavy quarkonium 
spectra . Using the location of the Schottky peaks, we may identify, 
using the intuition from the potential models, the average energy scales 
corresponding to the central and HF interactions. The main advantage 
here is that there is no complication arising from isospin assignments 
since $I=0$ unless they are exotic states. 
This simplicity is lost when we consider states which have 
one or more light quarks. Nevertheless, we may look at the data in this 
sector to gain further insights. In this subsection we look at the data 
on the specific heat in the open charm sector. The data used in the 
analysis is given in table \ref{tab3}.

\begin{table}[ht]
\centering
\captionsetup{justification=raggedright,
singlelinecheck=false
}
\begin{tabular}{|c|c|c|c|c|c|c|c|c|}\hline\hline
~$J=0$ States~&~$I$~& Mass &~$J=1$ States~&~$I$~& Mass &~$J=2$ States~&~$I$~& Mass \\
\hline\hline 
$D_s^{\pm}$           & 0            &1968.30 &$D_{s1}(2460)^{\pm}$  & 0            &2459.5  &$D_2^*(2460)$& $\frac{1}{2}$ &2463.453\\\hline
$D_{s0}^*(2317)^{\pm}$& 0            &2317.7  &$D_{s1}(2536)^{\pm}$  & 0            &2535.10 &             &               &\\\hline
$D$                   & $\frac{1}{2}$&1867.225&$D_{s1}^*(2700)^{\pm}$& 0            &2709    &             &               &\\\hline
$D_0^*(2400)^0$       & $\frac{1}{2}$&2318    &$D^*(2007)$           & $\frac{1}{2}$&2008.61 &             &               &\\\hline
$D_0^*(2400)^{\pm}$   & $\frac{1}{2}$&2403    &$D_1(2420)$           & $\frac{1}{2}$&2422.3  &             &               &\\\hline
$D_1(2420)^{\pm}$     & $\frac{1}{2}$&2539.4  &$D_1(2430)^0$         & $\frac{1}{2}$&2427    &             &               &\\\hline
\hline
\end{tabular}
\caption{Masses of open charm mesons given in MeV along with their total $J$ and $I$ }
\label{tab3}
\end{table}

\begin{figure}[ht]
\centering
\includegraphics[width=0.8\textwidth]{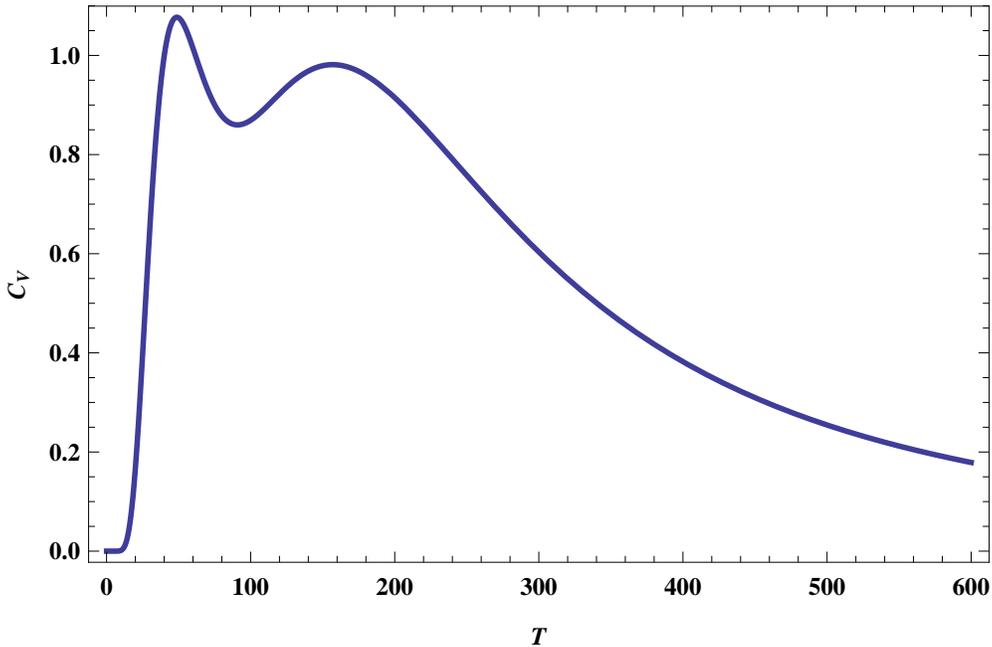}
\caption{$\text{C}_{\text{v}}$ vs $T$ plot for open charm mesons with 
all the states in Table.~\ref{tab3}}
\label{cmes}
\end{figure}

Fig.~\ref{cmes} shows the specific heat calculated using all the states 
in the open charm spectrum plotted as a function of $T$. We see two 
peaks at temperatures close to $T=50$ and $T=160$ in MeV units. As usual
the peak at higher $T$ represents the flavour independent part of the
potential. The scale of HF interactions is larger as it should be since
the splitting involves a light and a heavy quark. 

\begin{figure}[ht]
\centering
\subfloat[][]{\includegraphics[width=0.49\linewidth]{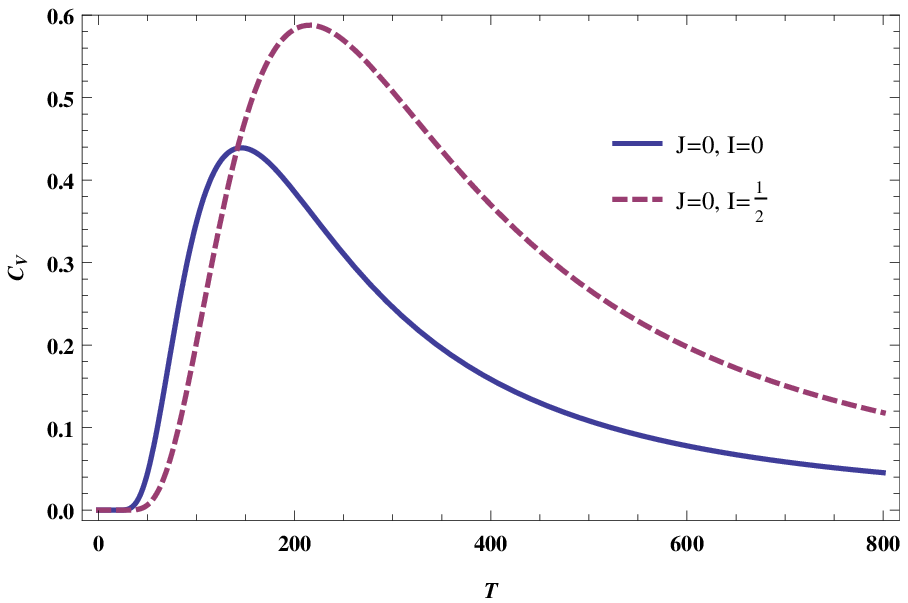}\label{cmesJ0}}\hfill
\subfloat[][]{\includegraphics[width=0.49\linewidth]{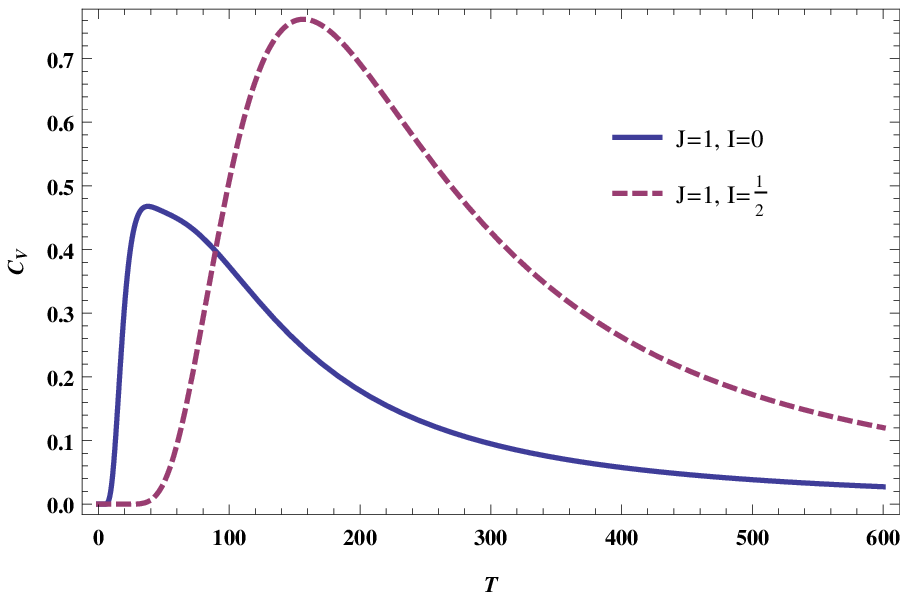}\label{cmesJ1}}
\caption{$\text{C}_{\text{v}}$ vs $T$ plot for open charm mesons
 (a) with $J=0$ states (column 3 of Table.~\ref{tab3}) 
 (b) with $J=1$ states (column 6 of Table.~\ref{tab3})
 separated according to the $I$ values}
\end{figure}

We may analyze the data further by separating them according to the 
spins. The $J=2$ sector cannot display any Schottky peak since 
it only has one state with $I=\frac{1}{2}$ and none with $I=0$.
 In Fig.~\ref{cmesJ0} we show the $J$=0 spectra. The two peaks 
above 150 MeV correspond to isospins $I=0$ and $I=1/2$. In the latter 
case the peak shifts to the right since there are more states as seen in 
the effect of truncation in the ideal case discussed in Sec. 2. 
The $J=1$ sector is shown in Fig.~\ref{cmesJ1} where again
we have shown the $I=0$ and $I=1/2$ cases separately. In the $I=1/2$ 
case one may clearly see the usual confinement peak close to $T=160$ MeV. 
The $I=0$ plot indicates a single peak with a shoulder at $40$ MeV. 

\begin{figure}[ht]
\centering
\includegraphics[width=0.8\textwidth]{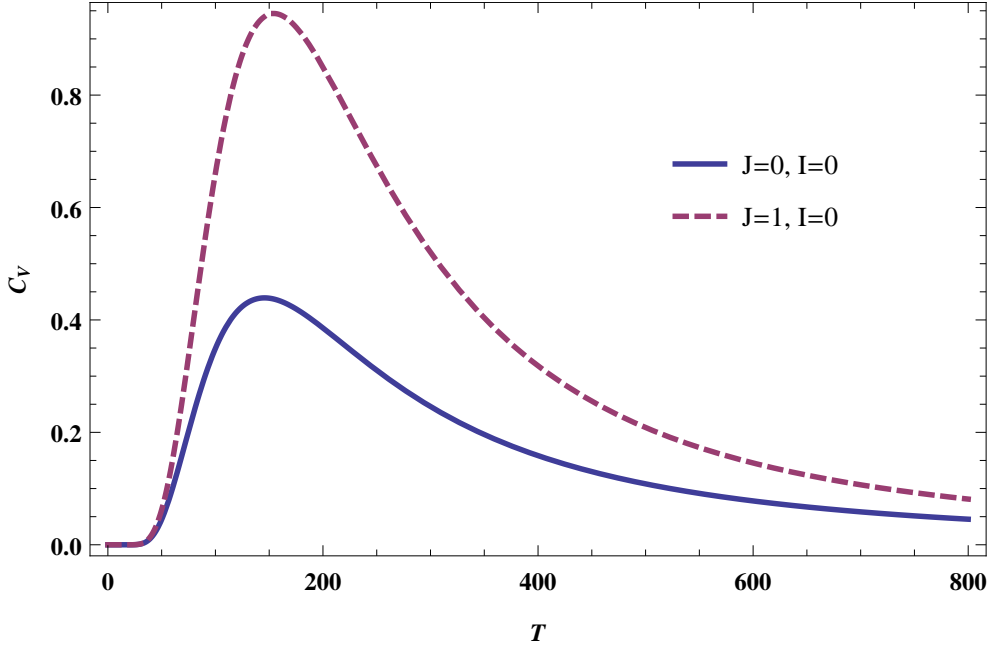}
\caption{$\text{C}_{\text{v}}$ vs $T$ plot for $I=0$ states grouped into $J=0$ and
$J=1$ after the assigning $J=1$ to $D_s^{*\pm}$ with mass 2112.1 MeV}
\label{I0mod2}
\end{figure}

PDG includes $D_s^{*\pm}$ with mass 2112.1 with $I=0$, but states that its 
$J$ is unmeasured. However, the corresponding decay modes of this state are 
consistent with $J=1$. Including this state in the $J=1$, $I=0$ set,  
immediately brings the corresponding peak to approximately the same 
position as the $J=0$, $I=0$ peak (Fig.~\ref{I0mod2}). Hence, this 
confirms that this state actually is indeed a $J$ triplet. 

\subsection{The spectra of open bottom states}

Unlike the open charm spectra, the spectra of open bottom states is even
more sparse as shown in Table~\ref{tab5}. 

\begin{figure}[ht]
\centering
\subfloat[][]{\includegraphics[width=0.49\linewidth]{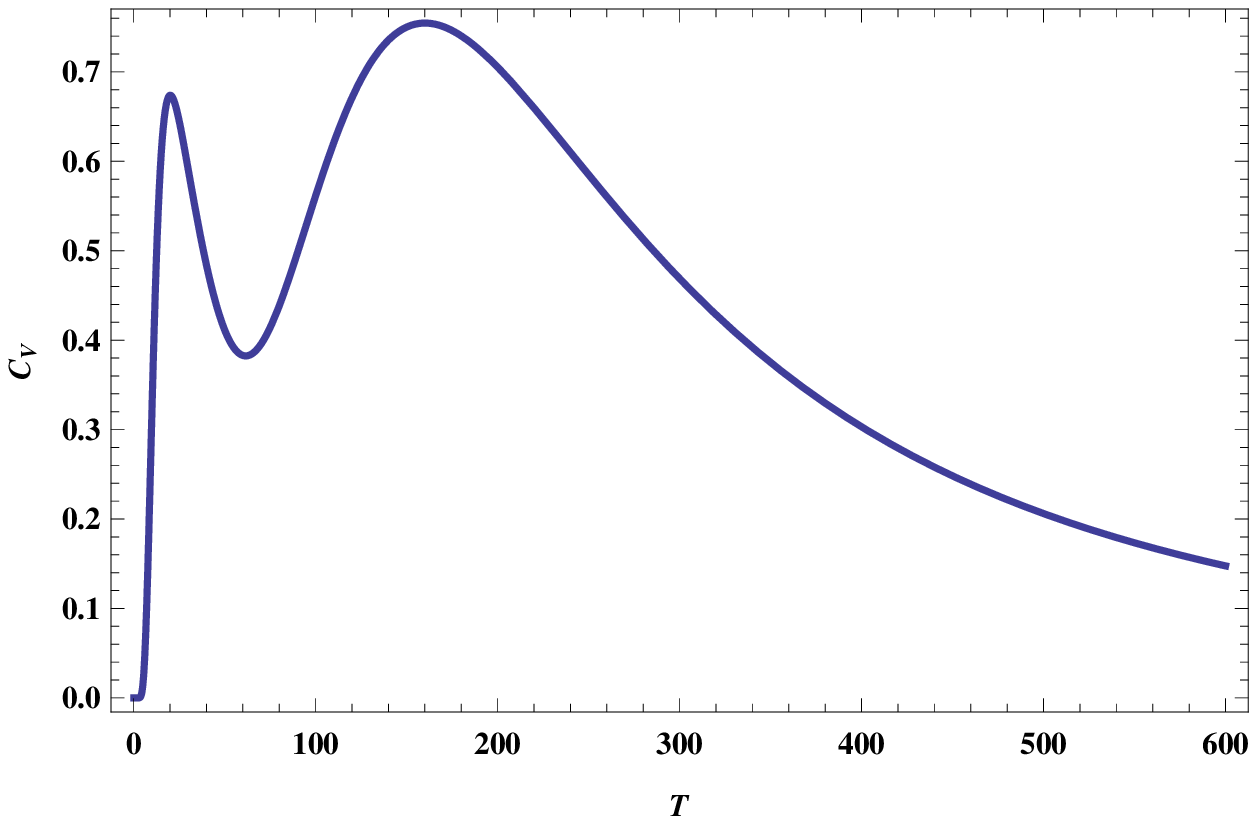}\label{bmes}}\hfill
\subfloat[][]{\includegraphics[width=0.49\linewidth]{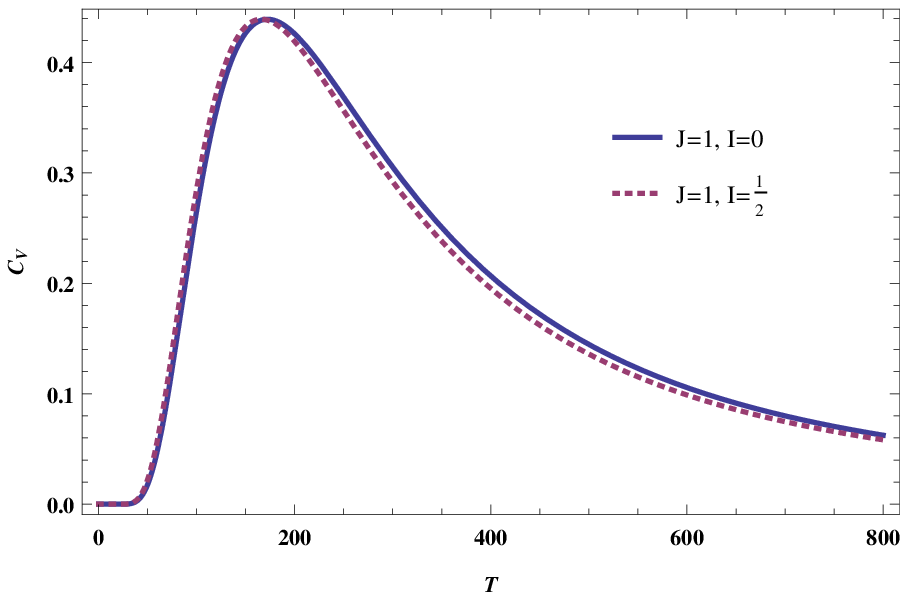}\label{bmesJ1}}
\caption{$\text{C}_{\text{v}}$ vs $T$ plot for open bottom mesons (a) with all the 
states given in Table.~\ref{tab5} (b) with $J=1$ states (column 6 of Table.~\ref{tab5})
grouped according to the $I$ values} 
\end{figure}

\begin{table}[ht]
\centering
\captionsetup{justification=raggedright,
singlelinecheck=false
}
\begin{tabular}{|c|c|c|c|c|c|c|c|c|}\hline\hline
~$J=0$ States~&~$I$~& Energy &~$J=1$ States~&~$I$~& Energy &~$J=2$ States~&~$I$~& Energy \\
\hline\hline 
$B_s^0$    & 0             &5366.77 &$B_s^*$         & 0             &5415.4  &$B_{s2}^*(5830)^0$& 0             &5839.96\\\hline
$B_c^{\pm}$& 0             &6275.6  &$B_{s1}(5830)^0$& 0             &5828.7  &$B_2^*(5747)^0$   & $\frac{1}{2}$ &5743   \\\hline
$B$        & $\frac{1}{2}$ &5279.42 &$B^*$           & $\frac{1}{2}$ &5325.2  &                  &               &       \\\hline
           &               &        &$B_1(5721)^0$   & $\frac{1}{2}$ &5723.5  &                  &               &       \\\hline
 \hline
\end{tabular}
\caption{Masses of open bottom mesons given in MeV along with their total $J$ and $I$}
\label{tab5}
\end{table}
Nevertheless some features are
already visible. For example when all the states are plotted together 
as shown in Fig.~\ref{bmes}, we discern the confinement as well as
the HF peak (at a reduced temperature), similar to the open charm scenario.

Separation into various $J$ and $I$ lists cannot be carried out here 
since there are only two states each in the $J=0$, $I=0$; $J=1$, $I=0$ 
and $J=1$, $I=\frac{1}{2}$ sectors. It is not possible to plot the 
$J=0$, $I=0$ states against $T$ due to different flavour contents. For 
example, 5366.77 MeV state is a $B_s$ (b and s quark) state whereas the 
6275.6 MeV state is a $B_c$ (b and c quark) state.

For completeness, we may however plot the $J=1$, $I=0$ and $J=1$, 
$I=\frac{1}{2}$ states each of which has only two states. We have showed 
the plot in Fig.~\ref{bmesJ1} where again the peaks coincide indicating
the confinement scale. More detailed analysis must wait for more data.

\section{Summary and conclusions}

We have presented a model independent analysis of the data on the meson 
spectra using the Schottky anomaly. Given a spectrum of states, the 
specific heat (or equivalently energy fluctuation) when plotted as a 
function of temperature displays peaks, known as Schottky peaks, 
corresponding to the different scales present in the interaction 
Hamiltonian which gives rise to such states. The peaks are well pronounced 
especially if the spectrum is truncated with very few orbitals as in the 
case of meson spectra. The corresponding temperature at which the Schottky 
peaks occur may be converted into energy scales relevant to the spectra 
on hand. This method is well known in other areas of Physics.

In this analysis temperature is simply treated as a parameter which is 
used to extract the scale and should not be confused with the 
thermodynamic temperature. The information so obtained is nothing new 
since it is already contained in the spectra, but it provides a new way of 
looking at the data and analyzing the same. This is especially useful if a 
given set of states, in the absence of any other information, contains 
states which may have their origin in different types of interaction 
Hamiltonians, for example the presence of exotics in the quarkonium 
states. Unlike the models of quarkonium states, no dynamical information 
may be extracted. But the intuition from the potential models or Lattice 
calculations may be used to gain more insight.

After explaining the salient features of the method through an ideal case,
we have analyzed the experimental data as listed in the latest edition of
PDG. In summary we have:
\begin{itemize}

\item The simplest to analyze are the quarkonium states, especially the 
charmonium states. Using this as a template we show how the two main 
scales corresponding to flavour independent confinement interaction and 
the HF interaction may be seen from the representation of the data through 
Schottky peaks. Typically the average confinement scale (corresponding to 
the radial potential in a model) results in a peak around $T=200$ MeV. If 
we are to interpret this information in terms of the actual scale of 
confinement we need to invoke a model. If there are only two to three 
orbitals involved (as is the case in most of the data) then a ball park 
estimate of the confinement scale is given approximately by $2.4T$. It 
varies very slowly with truncation of the orbitals as seen in the ideal 
case. The Schottky peak corresponding to HF interaction results in a peak 
around $T=40$ MeV which results in an average splitting of about 
90 MeV as given by potential models.

An analysis of the so called exotic states in the charmonium sector seems
to indicate that they are indeed unusual states due to the absence of the 
usual confinement peak, but exhibit a peak at a lower $T$, about $100$ MeV, 
corresponding to a lower $\textquoteleft$exotic confinement' scale at 
$\sim 240$ MeV.

\item Our conclusions in the case of bottomonium states are similar to the 
charmonium states. The HF interaction is much weaker here as it should be 
if one uses the intuition from potential models. These states also exhibit 
the confinement peak at around $200$ MeV. Interestingly, the so called 
exotics again display a peak at about $100$ MeV, similar to the exotics 
in the charmonium sector.

\item The analysis of open charm and open bottom states is more 
complicated but nevertheless we do get some insight into the scales 
involved.

\end{itemize}

Finally without going into details, we add a few comments on the light quark sector.
The light quark sector is the most complicated since here the 
intuition from the potential models are not as clean as in the case of the 
heavy quark sector. Furthermore, the HF splitting is comparable to the 
confinement scale. This results in a single broad peak which may be due to 
a combination of confinement and HF splitting. Both in non-strange and strange 
quark pseudo scalar bound states the confinement
scale shifts to a much larger value due to the complication arising from 
the masses of pion and K-meson. However, when $J\ne 0$ states are analyzed 
they display the confinement peak as witnessed in all other sectors.
This buttresses the well known problem with pions, that they are too light 
to be simple bound states of quark and an antiquark~\cite{compasspion}. This is also true,
to a lesser extent, with the pseudo scalar K-mesons around 495 MeV.

While we have analyzed the data on meson spectra here, we may analyze the 
baryon spectra also from this perspective. In some ways this later analysis 
is likely to provide complimentary information further simplified by the 
fact that only quarks are involved unless the exotics are considered.
This analysis is underway and will be published later~\cite{up}.

\acknowledgements 

We thank R. K. Bhaduri for many suggestions and careful reading of the 
manuscript. The ideas outlined in the paper originated during the
discussions in the HEP Journal Club. We thank all the members of JC
for their helpful comments, especially Srihari Gopalakrishna, D 
Indumathi and Rahul Sinha.

%   \bibliography{cite-paper}

% merlin.mbs apsrev4-1.bst 2010-07-25 4.21a (PWD, AO, DPC) hacked
% Control: key (0)
% Control: author (8) initials jnrlst
% Control: editor formatted (1) identically to author
% Control: production of article title (-1) disabled
% Control: page (0) single
% Control: year (1) truncated
% Control: production of eprint (0) enabled
%

\end{document}